\documentclass[]{elsarticle}

\usepackage{amsmath}
\usepackage{lineno}
\modulolinenumbers[5]

\biboptions{authoryear}

\usepackage{color}

\usepackage{array}
\usepackage{csquotes}
\usepackage{enumitem}
\usepackage{amssymb}
\usepackage[colorlinks,urlcolor=blue]{hyperref}
{\left\lbrace\begin{array}{@{}l@{}}}%
{\end{array}\right.}

\begin{document}

\begin{frontmatter}

\title{Snap buckling of bistable beams under combined mechanical and magnetic loading}


\author[mymainaddress]{Arefeh Abbasi}
\ead{arefeh.abbasi@epfl.ch}

\author[mymainaddress]{Tomohiko G. Sano \fnref{T}}
\ead{tomohiko.sano@keio.jp}
\fntext[T]{Current affiliation: Department of Mechanical Engineering, Keio University, Japan}

\author[mymainaddress]{Dong Yan}
\ead{dong.yan@epfl.ch}

\author[mymainaddress]{Pedro M. Reis\corref{mycorrespondingauthor}}
\cortext[mycorrespondingauthor]{Corresponding author}
\ead{pedro.reis@epfl.ch}

\address[mymainaddress]{\'{E}cole Polytechnique F\'{e}d\'{e}rale de Lausanne (EPFL), Flexible Structures Laboratory, CH-1015 Lausanne, Switzerland\\
    }

\begin{abstract}
We investigate the mechanics of bistable, hard-magnetic, elastic beams, combining experiments, finite element modeling (FEM), and a reduced-order theory. The beam is made of a hard magneto-rheological elastomer, comprising two segments with antiparallel magnetization along the centerline, and is set into a bistable curved configuration by imposing an end-to-end shortening. Reversible snapping is possible between these two stable states. First, we experimentally characterize the critical field strength for the onset of snapping, at different levels of end-to-end shortening. Second, we perform 3D FEM simulations using the Riks method to analyze high-order deformation modes during snapping. Third, we develop a reduced-order centerline-based beam theory to rationalize the observed magneto-elastic response. The theory and simulations are validated against experiments, with an excellent quantitative agreement. Finally, we consider the case of combined magnetic and mechanical-indentation loading, examining how the applied field affects the bistability and quantifying the maximum load-bearing capacity. Our work provides a set of predictive tools for the rational design of one-dimensional, bistable, magneto-elastic structural elements.
\end{abstract}


\end{frontmatter}

\newpage
\section{Introduction}
\label{sec:Introduction}
Bistable structures are central in the design  of many functional devices~\cite{cao2021bistable, wang2017harnessing, ghosh2016multistability,harne2013review, sun2019stochastic}, whose internal energy comprises two minima, separated by a maximum, representing the barrier to a fast transition between two stable states. This snap-through instability can be exploited to cause relatively large displacements, or rotations, with low work for actuation, offering potential applications in several engineering domains, including micro-electromechanical systems~\cite{vangbo_lateral_1998,saif_tunable_2000}, robotics~\cite{chen_harnessing_2018,rothemund_soft_2018}, energy harvesting~\cite{stanton_nonlinear_2010, clingman_development_2017}, actuators~\cite{crivaro_bistable_2016, hou2018magneto}, origami structures~\cite{treml_origami_2018,faber_bioinspired_2018}, and deployment mechanisms~\cite{chen_integrated_2017}. Bistable beams can be classified into two categories~\cite{liu2015multistable}: (i) pre-shaped, which do not possess residual stresses, and (ii) pre-compressed, which are stressed post-production to exhibit the first buckling-mode configuration~\cite{fang1994post}. The latter have gained much attention due to their manufacturability and versatility~\cite{jeon2010snap, hou2018magneto, cleary2015modeling}. Next, we provide an overview of recent research centered on pre-compressed beams, the type central to our study.

Many past studies on pre-compressed beams have focused on their snap-through characteristics~\cite{fang1994post, vangbo_lateral_1998, vangbo1998analytical, saif_tunable_2000, yan2019analytical}: the critical load and displacement, and the travel distance from the first stable state to the new configuration. These features can be set by design parameters; \textit{e.g.}, the beam geometry, end-to-end shortening~\cite{vangbo_lateral_1998}, actuation loading and position~\cite{zhao2008post,cazottes2009bistable,camescasse2013bistable,camescasse2014bistable,harvey2015coexisting, cazzolli2019snapping,zhang2020configurations, wan2020tunable}, and boundary conditions~\cite{plaut2015snap,camescasse2013bistable,camescasse2014bistable,cleary2015modeling}. A recently emerging trend in the field of bistable beams is the usage of active materials with external stimuli to control the stability during snap-through by inducing local strains from temperature gradients, swelling, or electric/magnetic fields. For example, electrostatic~\cite{krylov2011bistability,medina2018bistability}, piezoelectric~\cite{maurini2007distributed}, and magnetic~\cite{ramachandran2016elastic,amor2020snap} actuation have all been used to control the bistability. 

More specifically, there has been a burgeoning interest in magneto-rheological elastomers (MREs), comprising magnetically permeable micron-sized particles embedded into a polymeric matrix, with a mechanical response that can be tuned under an external magnetic field~\cite{carlson2000mr, danas2012experiments}. Structures made out of MERs offer opportunities for fast, reversible, and remotely controlled shape-shifting behavior~\cite{gray2014review, li2014state,bica2014hybrid, hines2017soft,Kim_Nature2018, hu2018small, wu2020multifunctional, chen2021reprogrammable,wang2021evolutionary, kim2022magnetic}. Based on the magnetic response of the embedded particles to applied magnetic fields, MERs are classified into two groups, soft-MREs (s-MERs) or hard-MERs (h-MERs), each of which is discussed next.

In s-MREs, the magnetization of the particles changes in response to an external field~\cite{miyazaki2012physics} and, when embedded in a compliant elastomeric matrix, the particles tend to form chains along the field direction~\cite{ginder2002magnetostrictive,danas2012experiments}. The displacements of the constituent micro-scale particles within the matrix result in macro-scale deformations and changes in the elastic properties of the composite~\cite{rigbi1983response,ginder1999magnetorheological,danas2012experiments}. To actuate the large deflection of structures made of s-MREs, a field gradient is applied to generate magnetic forces. For example, the remote magnetic actuation of bistable spherical caps made of s-MREs has been investigated~\cite{loukaides2014magnetic, seffen2016eversion}.

In h-MREs, the focus of the present study, the particles possess a high coercivity to resist demagnetization by external fields upon field saturation~\cite{miyazaki2012physics, wu2020multifunctional} and their remnant magnetization can be retained during actuation. In particular, flexible slender structures made of h-MREs are capable of significant shape changes, driven by the magnetic body torques induced by the interaction between the intrinsic magnetization of the material and the applied field~\cite{lum2016shape,Kim_Nature2018}. The magnetization profile of h-MRE structures can be designed by the local orientation of the magnetized particles to generate complex 3D-shape transformations and optimize the shape-shifting modes for specific applications~\cite{lum2016shape,Kim_Nature2018, ciambella2020form, chen2020complex, chen2020mechanics, wang2021evolutionary, alapan2020reprogrammable}.

Due to the elasticity-magnetism coupling, together with the underlying geometric nonlinearities, modeling the mechanical behavior of hard-magnetic soft structures is challenging but there have recent advances in this direction.   A continuum theory has been developed~\cite{zhao2019mechanics} for the finite deformation of 3D (bulk) h-MREs through a nonlinear magneto-mechanical constitutive law. In this framework, the Helmholtz free energy density includes elastic (neo-Hookean) and magneto-elastic terms. A simulation framework by the same authors using FEM was also developed. Subsequently a full-field 3D continuum model for h-MREs was proposed~\cite{mukherjee2021explicit}, also incorporating magnetic dissipation, particle-particle interactions, and the surrounding air effects. They validated their model by performing microscopic homogenization simulations applied to macroscopic boundary value problems. Based on the 3D continuum model, and using dimensional reduction, theories for inextensible, hard-magnetic elastica was derived and validated against experiments, under either a uniform magnetic field~\cite{wang2021evolutionary, lum2016shape} or a field with constant gradient~\cite{yan_comprehensive_2021}. A similar dimensional reduction approach was applied to model the 3D deformation of hard-magnetic rods under uniform and gradient magnetic fields~\cite{sano_kirchhoff-like_2022, sano-dipole-2022}. Considering the extensibility of the centerline, a geometrically exact beam model under uniform fields was developed to predict the deformation of cantilever beams~\cite{chen2020complex, chen2020mechanics, chen2020theoretical}, albeit finding negligible differences with the inextensible model. A similar strategy based on dimensional reduction has employed to capture the behavior of magnetic thin plates~\cite{yan2022reduced}, and predict the axisymmetric deformation of pressurized hard magnetic shells~\cite{yan2021magneto}. This 1D shell model was later generalized in a 3D configuration~\cite{pezzulla2022geometrically}. 

Even if there have been several studies on modeling the deformation of magneto-active structures, their instability and, more specifically, the snap-through phenomenon under magnetic actuation remains an ongoing research topic. Important questions to address include how bistable systems transit between stable configurations under a magnetic field and the contribution of various buckling modes and energy levels to this transition. Additionally, theoretical and computational tools are needed to predict the critical conditions and snap-through response of magneto-active structures. Such developments would be valuable for the predictive and rational design of bistable magneto-elastic systems.

Here, we study of the snap-through elastic bistable beams under magnetic actuation, combining theory, FEM, and  experiments. First, we demonstrate that snap-buckling can be triggered in the presence of an external uniform magnetic field. We quantify how the critical field strength required for buckling depends on the imposed end-to-end shortening (setting the pre-buckled configuration), the beam geometry, and the material and magnetization properties. A centerline-based theory is developed to rationalize the trade-offs between the various loading and geometric parameters, predicting the conditions for the onset of snapping. In parallel, we adapt the finite element method (FEM) for 3D h-MREs proposed by in Ref.~\cite{zhao2019mechanics} to make it amenable to Riks (arc-length) analysis. With this enhancement, it is possible to track the stable and unstable branches of the load-displacement curve during snapping. We also probe the beam's load-bearing capacity when the external loading combines a constant magnetic field and mechanical indentation. 

Our paper is organized as follows. In Section~\ref{sec:problemdef}, we define the problem at hand. In Section~\ref{sec:experiments}, we present the experimental method to fabricate the h-MRE beam specimens and describe the experimental protocol for snap-buckling tests. The FEM simulations using the Riks method are detailed in Section~\ref{sec:numerics}. In Section~\ref{sec:theory}, we derive a 1D reduced-order model for bistable magnetic beams. Then, in Sections~\ref{sec:beam_mech}-\ref{sec:Beam_mechmag} we report the experimental results and their comparisons with the theoretical and FEM predictions. Finally, our main contributions and an outlook for future work are summarized and discussed in Section~\ref{sec:conclusion}.
\section{Definition of the problem}
\label{sec:problemdef}

\begin{figure}[h!]
    \centering
    \includegraphics[width=0.75\columnwidth]{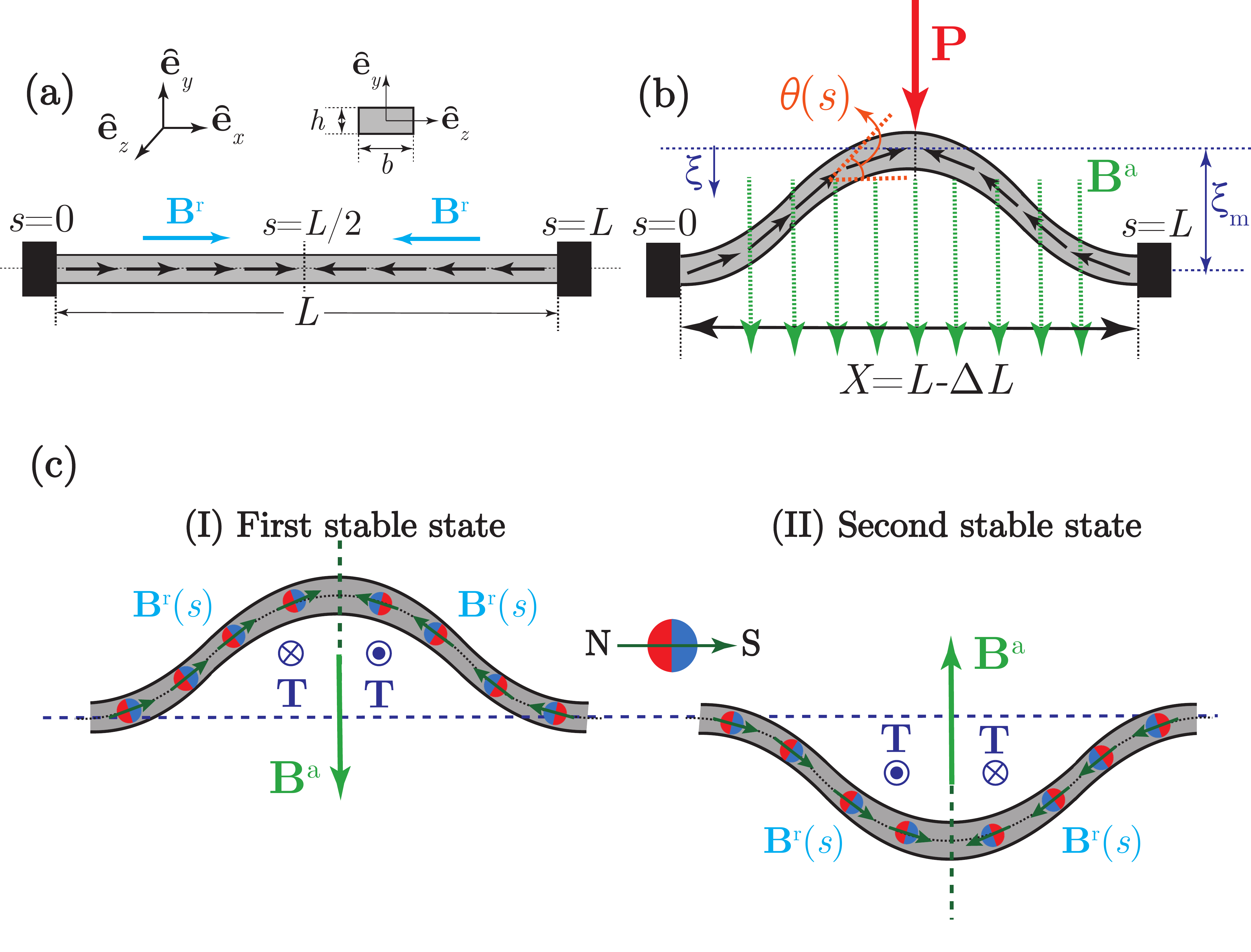}
    \caption{Definition of the problem. (a) Schematic diagram of the undeformed configuration of a naturally straight beam of initial length $L$, thickness $h$, and width $b$. The beam is composed of two segments with antiparallel magnetization, $\mathbf{B}^\mathrm{r}$, along the centerline according to Eq.~(\ref{eq:magnetization}). (b) The beam is first pre-loaded by imposing a dimensional end-to-end shortening, $\Delta L$, thereby deforming to a curved, bistable configuration, and then made to snap between the two stable configurations under an applied uniform magnetic field, $\mathbf{B}^\mathrm{a}$, and/or an indentation force, $\mathbf{P}$. (c) The beam exhibits two stable equilibrium states; upon the application an external magnetic field, the generated torques can switch the beam between configurations (I) and (II).}
    \label{fig:definition}
\end{figure}

We seek to investigate the snap buckling of a bistable magneto-active beam under magnetic loading, which may also be combined with a mechanical (point) load. We consider a hard-magnetic, thin, elastic beam of length $L$ and rectangular cross-section of width $b$ and thickness $h$ (Fig.~\ref{fig:definition}a). The beam is made of an isotropic and homogeneous h-MRE material has Young's modulus, $E$, and Poisson's ratio, $\nu$. The configuration of the beam is described using the Cartesian basis vectors $(\mathbf{\hat{e}}_x,\,\mathbf{\hat{e}}_y,\,\mathbf{\hat{e}}_z)$, aligned, respectively, to the length, thickness, and width directions of the originally straight beam (Fig.~\ref{fig:definition}a). The beam is parameterized using the arc length coordinate, $0 \leq s \leq L$, along its centerline.

The left and right halves of the beam are magnetized permanently in opposite directions, parallel and anti-parallel to $\mathbf{\hat{e}}_x$, respectively, with the absolute residual magnetic flux density of $B^\mathrm{r}$ (Fig.~\ref{fig:definition}a). Given the slenderness of the beam, the residual magnetic flux density is assumed constant across the cross-section but varies along the arc length direction as:
\begin{equation}
\mathbf{B}^{\mathrm{r}}(s)=-B^{\mathrm{r}} \mathrm{sgn}\left(s-\frac{L}{2}\right) \mathbf{\hat{e}}_{x}.
\label{eq:magnetization}
\end{equation}
Having also compared this magnetization profile with the uniform case, $\mathbf{B}^{\mathrm{r}}=-B^{\mathrm{r}}\mathbf{\hat{e}}_{x}$, we found that the profile in Eq.~(\ref{eq:magnetization}) is more effective in inducing snap buckling. Even if we recognize that the present choice is ad hoc, it works and it is simple; we leave a more systematic exploration of other magnetization profiles for future work.  The magnetic loading is exerted on the beam by the application an external magnetic field, $\mathbf{B}^\mathrm{a}$. Due to the profile of the residual magnetic flux density vector, $\mathbf{B}^\mathrm{r}$, with respect to the direction of the applied magnetic field, $\mathbf{B}^\mathrm{a}$, we will demonstrate that the proposed configuration can induce snap-through buckling of the beam. 

The magnetized beam is naturally straight in its initial configuration, with the two ends clamped at $s=0$ and $s=L$ (Fig.~\ref{fig:definition}a). To form a buckled (bistable) beam, we then impose a dimensional end-to-end shortening, $\Delta L$, by translating the  end at $s=L$ (Fig.~\ref{fig:definition}b), such that the projected length of the beam becomes $X=L-\Delta L$. The (dimensionless) end-to-end shortening is then defined as $\epsilon=\Delta L/L$. In the initial curved configuration set by $\epsilon$ with the first-buckling-mode shape, considering $w$ as the displacement of the beam in the $\mathbf{\hat{e}}_y$ direction, the vertical rise of the beam's mid-span along $\mathbf{\hat{e}}_y$, is denoted by $\xi_\mathrm{m}$. Subsequently, the mid-span displacement $w(s=L/2)=\xi\neq \xi_\mathrm{m}$  will vary  when external loads are applied. The deformed configuration of the beam is described by the angle, $\theta (s)$, between the tangent of the centerline and $\mathbf{\hat{e}}_x$ (Fig.~\ref{fig:definition}b), with clamped boundaries; $\theta(0)=\theta(L)=0$.

Once pre-loaded into a curved configuration, the beam can undergo a snap-through instability by the application of either an indentation force, $\mathbf{P}$, or a uniform magnetic field, $\mathbf{B}^\mathrm{a}$, or the combination of the two (Fig.~\ref{fig:definition}b). Specifically, under an external magnetic field, a magnetic body torque, $\mathbf{T}$, results from the resistance to rotation of the vector of the residual magnetic flux density, $\mathbf{B}^\mathrm{r}$, which tends to align $\mathbf{B}^\mathrm{r}$ with the applied magnetic field, $\mathbf{B}^\mathrm{a}$ (Fig.~\ref{fig:definition}c). Hence, snap-buckling is initiated primarily driven by the magnetic body torque~\cite{zhao2019mechanics}
\begin{equation}
\mathbf{T}=\frac{1}{\mu_{0}} \mathbf{B}^\mathrm{r} \times \mathbf{B}^{\mathrm{a}},
\end{equation}
where $\mu_\mathrm{0}$ is the vacuum permeability. After switching from one stable state (Fig.~\ref{sec:problemdef}cI) to another (Fig.~\ref{sec:problemdef}cII), and removing the exterior magnetic field, the beam stays in the second stable position. The process can be reversed by applying a magnetic field with opposite polarity. In the case of snap buckling under simultaneous mechanical and magnetic loading, the beam is first loaded under a prescribed value of the uniform magnetic field and then indented by a concentrated load applied, $\mathbf{P}=-P\mathbf{\hat{e}}_y$ at its mid-span (Fig.~\ref{fig:definition}b). 

Whereas many previous studies have addressed the critical conditions for the classic problem of snap buckling of elastic bistable beams \cite{vangbo_lateral_1998,das_pull-and_2009,saif_tunable_2000,Pandey_2014_EPL,camescasse2013bistable,camescasse2014bistable,Sano:2018_PRE,amor2020snap}, in this study, we investigate the conditions for snapping of a bistable, hard-magnetic beam under the combined influence of magnetic and mechanical (indentation) loads.

\section{Experimental methods}
\label{sec:experiments}

This section presents the experimental methodology, first, describing the fabrication of the beam specimens (Section~\ref{sec:fabrication}) and, then, detailing the experimental apparatus (Section~\ref{sec:apparatus}). Finally, we describe the experimental protocols, parameters, and procedures (Section~\ref{sec:procol}).

\subsection{Fabrication of the beam specimens}
\label{sec:fabrication}

The beam specimens were fabricated using a casting protocol adopted from our recent work~\cite{yan_comprehensive_2021}. The main modification from our previous work is the process we use to magnetize the specimens, with two symmetric regions of antiparallel magnetization. This specific magnetization profile was chosen to facilitate snapping under magnetic loading. Still, for completeness, we provide an overview of the full fabrication protocol.

The beam specimens were cast with an h-MRE, prepared by mixing NdPrFeB particles (average diameter of  $5 \mu\,\mathrm{m}$, mass density of $\rho_\mathrm{mag}=7.61\,\mathrm{g/cm}^3$, MQFP-15-7-20065-089, Magnequench) with Vinylpolysiloxane (VPS) polymer (VPS-22, mass density of $\rho_\mathrm{vps}=1.16 \mathrm{g/cm}^3$, Elite Double, Zhermack). The fraction of NdPrFeB particles in the h-MRE was 50.0$\%$ in mass ($c_v=13.2\%$ volume fraction). The mixed solution was injected into the a sandwich mold using a syringe to cast a straight beam. Upon curing of the h-MRE, the average Young's modulus was $E=1.16 \pm 0.04\,$MPa, the  density was $\rho=2.01 \pm 0.05$ g/cm$^3$, and the Poisson's ratio was assumed to be $\nu \approx 0.5$ (near incompressibility).

\begin{figure}[h!]
    \centering
    \includegraphics[width=0.75\columnwidth]{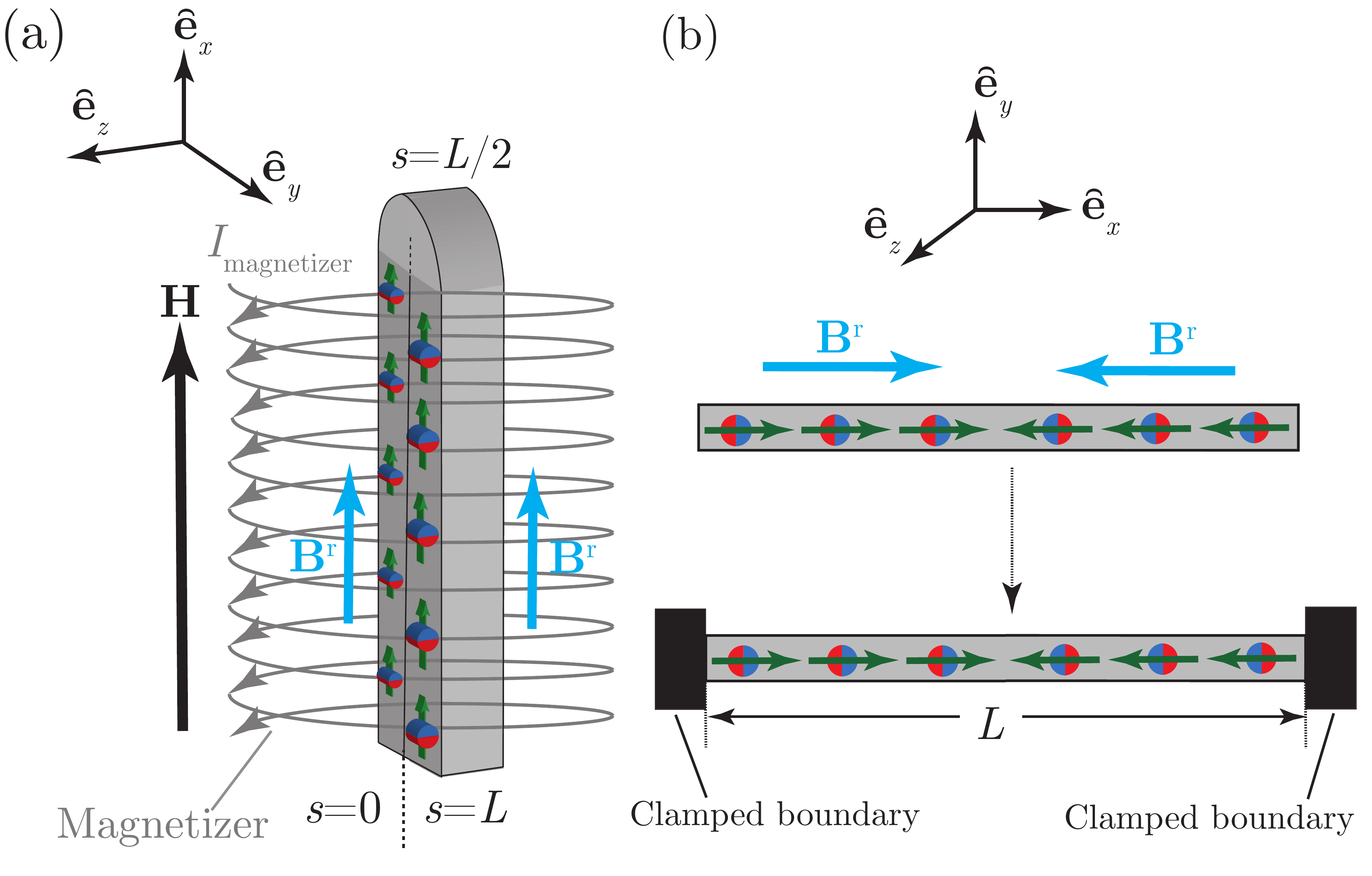}
    \caption{Magnetization of the beam specimen. (a) The beam was magnetized while folded at the mid-span, $s=L/2$, inside a pulse magnetizer, which generates a strong axial magnetic field, $\mathbf{H}$. (b) After unfolding, the magnetized specimen exhibited the antiparallel residual magnetic flux density of $\mathbf{B}^\mathrm{r}$ described in Eq.~(\ref{eq:magnetization}). Two non-magnetic cubic blocks attached to the extremities ensured clamped boundary conditions.}
    \label{fig:Exp}
\end{figure}

To achieve the desired magnetization profile, we folded the beam at mid-span (Fig.~\ref{fig:Exp}a) and placed it inside an impulse magnetizer (IM-K-010020-A, flux density $\approx$ 4.4 T, Magnet-Physik Dr. Steingroever GmbH). The magnetization of the embedded particles became permanently aligned to the direction of the field  generated by the magnetizer (Fig.~\ref{fig:Exp}a). Due to this folded configuration, after unfolding each half of the beam developed antiparallel magnetization (Fig.~\ref{fig:Exp}b), with the residual magnetic flux density, $\mathbf{B}^\mathrm{r}$, described in Eq.~(\ref{eq:magnetization}).  Assuming a uniform dispersion of the particles in the polymer matrix, and no re-arrangements during magnetization, the composite can be considered as a homogeneous continuum solid with a uniform magnetization on each half, whose magnitude was computed as the volume-average of the total magnetic moment of the individual particles, $M=94.1\,$kA/m.

After magnetization, two non-magnetic cubes of pure VPS ($8\times15\times15\,\textrm{mm}^3$) were mounted onto each of the beam extremities to set clamped boundary conditions (Fig.~\ref{fig:Exp}b). Finally, the end-to-end shortening, $\epsilon=\Delta L/L$, was imposed on the originally straight beam using an acrylic sample holder, exciting in the first buckling mode, with bistability, shown schematically in Fig.~\ref{fig:definition}.

\subsection{Experimental apparatus}
\label{sec:apparatus}

\begin{figure}[h!]
    \centering
    \includegraphics[width=0.8\columnwidth]{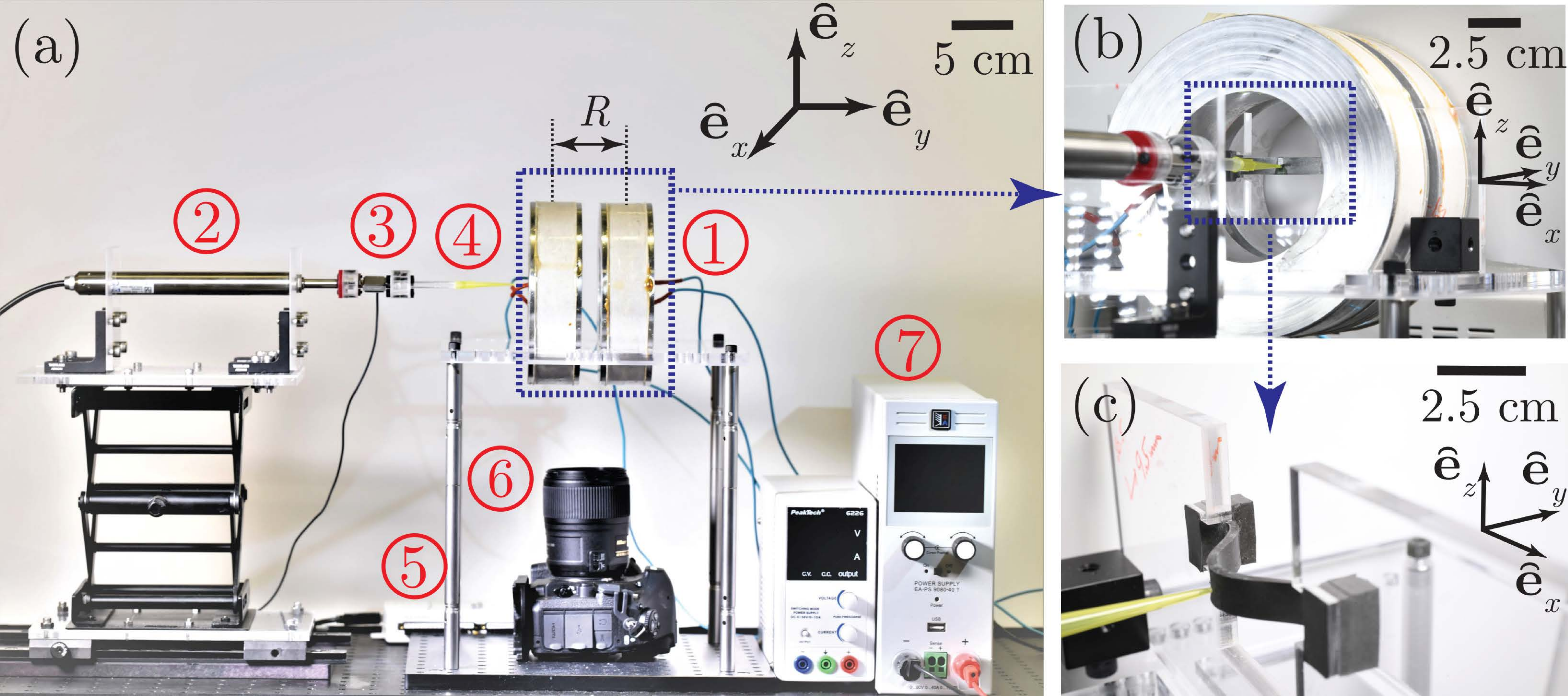}
    \caption{Photographs of the experimental apparatus. (a) A magnetic beam is positioned in between a set of Helmholtz coils (1) and loaded by an indenter (4). This indenter is attached to a motorized linear actuator (2), and the reaction force is monitored by a force sensor (3). The force-displacement data is acquired using a LabVIEW data acquisition card (5), and a camera (6) is used to image the beam profile. The coils are driven by the current output of a DC power supplier (7). (b) Zoomed view of the pair of Helmholtz coils. (c) Representative beam specimen positioned inside the coils.}
    \label{fig:setup}
\end{figure}

With the originally straight beam set in a curved (bistable) configuration, the experiments involved loading the specimen magnetically, or mechanically, or both, using the apparatus shown in Fig.~\ref{fig:setup}a. Gravitational effects were minimized by placing the beam with the deflection direction, $\hat{\mathbf{e}}_y$, perpendicular to gravity, $-g \hat{\mathbf{e}}_z$. A digital camera was set underneath the coils for imaging (Fig.~\ref{fig:setup}a6). 

For the magnetic loading tests, we used a pair of identical coaxial coils (different from the impulse magnetizer mentioned above), in a Helmholtz configuration, which generated a steady axial symmetric magnetic flux density, $\mathbf{B}^\mathrm{a} (x, y, z)$~\cite{yan2021magneto}. The coils were  connected in series  and separated axially by a distance equal to the mean radius of each coil ($R= 59.5\,$mm). In this configuration, the current was made to flow through both coils in the same direction to generate a uniform magnetic field in their central region (Fig.~\ref{fig:setup}a),
\begin{equation}
    \mathbf{B}^\mathrm{a}=B^\mathrm{a} \hat{\mathbf{e}}_y.
\end{equation}
Each coil was manufactured by winding an aluminum circular spool with an enameled copper wire (Repelec Moteurs S.A.). The dimensions of the coils were 86\,mm for the inner diameter, 152\,mm for the outer diameter, and 43\,mm in height. A DC power supply powered the coils, providing a maximum power of 1.5kW (EA-PSI 9200-25T, EA-Elektro-Automatik GmbH). The magnitude of the magnetic field, $B^\mathrm{a}$, was varied by adjusting the current output (0-25\,A) from the power supplier.  

For the mechanical load tests, we indented the beam specimen using a custom-built apparatus and measured the force-displacement curves. The indentation device comprised two parts: a high-resolution linear actuator (L-220.50DG, PI, Germany) driven by a 1-axis DC motor controller (C-863 Mercury Servo Controller, PI, Germany, Fig.~\ref{fig:setup}a2) to impose the displacement, and a force sensor (LRM200, 5lb, JR S-beam load cell, Futek, CA, USA, Fig.~\ref{fig:setup}a3) to measure the reaction force at an acquisition rate of 1 kHz. The indenter was a plastic (non-magnetic) needle cap (plastic taper tip Luer Lock 20GA$\times$1/4$''$ Vita needle, MA), chosen to avoid any magnetic field distortions. This indenter assumed rigid compared to the beam specimen was attached to the tip of linear actuator, as shown in Fig.~\ref{fig:setup}a4. The tip of the indenter was glued to the beam at mid-span using VPS solution, thereby restraining rotation and translation at the point of contact. This attachment enabled the acquisition of the complete load-displacement during indentation, including both stable and unstable paths.

\subsection{Experimental protocols}
\label{sec:procol}
To investigate the snap buckling of bistable hard magnetic beams, we  performed three sets of experiments with different loading conditions (i) mechanical point load, (ii) magnetic load, and (iii) combined mechanical and magnetic load using, respectively, the indentation apparatus, the coils, or both. The corresponding results from these experiments will be presented in Sections~\ref{sec:beam_mech},~\ref{sec:beam_mag}, and~\ref{sec:Beam_mechmag}, respectively. Next, we detail the configurations of the fabricated specimens, the range of parameters explored, and the experimental protocols.  

We tested three separate, but otherwise identical, beams (length $L = 60.00 \pm 0.10\,$mm, width $b = 8.00 \pm 0.04\,$mm, and thickness $h=2.00 \pm 0.06\,$mm) to examine the experimental reproducibility and uncertainty. Throughout the experiments, the slenderness ratio was kept constant, $k=L/h=30$. The end-to-end shortening was varied in the range $0 \leq \epsilon=\Delta L/L \leq 0.6$. Next, we describe each of the experimental tests, under the different loading conditions.


(i) \textit{Snap-through under mechanical point loading}: In order to capture the stable and unstable portions of the loading path, pure mechanical indentation, in the absence of a magnetic field was applied along $\hat{\mathrm{\mathbf{e}}}_y$, with the indenter glued to the beam at mid-span ($s=L/2$), and at the constant velocity of 0.02\,mm/s to ensured quasi-static conditions. For each level of $\epsilon$, the mid-span displacement varied in the range $0 \leq \xi \leq 2\xi_\mathrm{m}$. 

(ii) \textit{Snap-through under magnetic loading}: In a second set of experiments, the beam specimen was placed  within the region of uniform magnetic field generated by the Helmholtz coils~\cite{yan_comprehensive_2021}. Two different protocols were followed to measure (ii.a) the critical magnetic field for snapping, $B^\mathrm{a}_\mathrm{cr}$, and (ii.b) the full load-displacement response, $B^\mathrm{a}(\xi)$, as detailed next. 
(ii.a) To measure $B^\mathrm{a}_\mathrm{cr}$, we gradually increased the magnitude of the applied magnetic flux density (by increasing the current, $I$, in the coils; steps of 0.05\,A and 10\,s) until snap-through occurred. Assuming the snap-through phenomena is nearly instantaneous, and the waiting time between each two step is larger than the viscous relaxation time~\cite{gomez2017critical,gomez2019dynamics}, we neglected dynamic effects and measured the critical snapping magnetic field at the snapping step. (ii.b) By adapting the above experimental procedure, we characterized the full bistable response, capturing the stable and unstable paths. First, prior to magnetic loading, the indentation was performed under displacement-control conditions at the speed of 0.02\,mm/s (along $\hat{\mathrm{\mathbf{e}}}_y$). The indenter was then stopped at each step of 0.2\,mm, and the magnetic field was increased from zero, in the $-\hat{\mathrm{\mathbf{e}}}_y$ direction, to balance the indentation force, until a zero-force was measured by the load cell. Assuming equilibrium of the specimen and the quasi-static experimental conditions, the measured applied magnetic field required for zero-force was ensured to lie on the equilibrium curve, $B^\mathrm{a}(\xi)$. 

(iii) \textit{Snap-through under combined mechanical and magnetic load}: In a third set of experiments, we investigated the effect of magnetic loading on the snap-through response of the beam under simultaneous mechanical indentation. In each experimental run, the beam was first loaded under a steady magnetic field and then indented following the same protocol as in (i). We repeated the experiment at eleven different levels of magnetic field strength, in the range $-4.5\,\mathrm{mT}\leq\mathbf{B}^\mathrm{a}\leq 53.4\,\mathrm{mT}$, in steps of 6\,mT, ensuring that $B^\mathrm{a}_\mathrm{cr} \leq B^\mathrm{a}$. From the measured curves of indentation force versus displacement, $P(\xi)$, we characterized the stability of the beam for these combined loading conditions. 

\section{Numerical simulations using FEM}
\label{sec:numerics}

In parallel to the experiments, we performed 3D FEM simulations using an existing user-defined element~\cite{zhao2019mechanics} in the commercial software package ABAQUS/Standard 6.14. As detailed in Section~\ref{sec:FEM_riks}, we have modified this user element to enable Riks analysis on hard-magnetic structures. The Riks algorithm allows for the solution of the equilibrium equation of a structure by prescribing the arc-length of its loading path, so as to track both stable and unstable equilibrium states. We use this technique to study the snapping behavior of our magnetic beam subjected to a uniform magnetic field. The simulation procedure is, then, detailed in Section~\ref{sec:FEM_procedure}.
\subsection{User element for Riks analysis}
\label{sec:FEM_riks}

Our FEM approach is based on an existing continuum theory of ideal hard-magnetic soft materials~\cite{zhao2019mechanics} with a permanent magnetization independent of external magnetic fields. In this theory, the effect of an applied magnetic field on a magnetized, deformable body is considered through a potential (density) as a function of the deformation gradient ($\mathbf{F}$), the external field flux density ($\mathbf{B}^\mathrm{a}$) and the magnetization of the material (${\mu_{0}}^{-1}\mathbf{B}^\mathrm{r}$):
\begin{equation}
\hat{U}^\mathrm{m}={\mu_{0}}^{-1}\mathbf{F}\mathbf{B}^\mathrm{r}\cdot \mathbf{B}^{\mathrm {a}}\,.
\label{Um_3D_beam}
\end{equation}
This magnetic potential is added to the total energy of the system. Under this description, distributed magnetic torques imposed by the applied field result in an asymmetric part of the Cauchy stress. The field produced by the magnetic body and the induced self-interactions are neglected. This theory has been previously implemented in the commercial FEM software package ABAQUS through a user-defined 8-node solid element~\cite{zhao2019mechanics}, while assuming that the elastic behavior of the material is assumed to be neo-Hookean. 

To capture the unstable equilibrium path of the snapping beam under magnetic actuation, we had to adapt this previously developed user element to make it compatible with the Riks analysis in ABAQUS. In the Riks analysis, the magnitude of external loads, which is usually prescribed during a simulation, is considered as an unknown and solved simultaneously with displacements from equilibrium. Alternatively, the `arc length of the static equilibrium path of a system in the load-displacement space is imposed to control the progress of the simulation. In order to implement the Riks analysis in the presence of a uniform magnetic field, we define its magnitude as a loading parameter using the keyword \texttt{*DLOAD} in ABAQUS, rather than a field variable in the previous work~\cite{zhao2019mechanics}. As such, the field strength can be taken in account as an unknown in the solution domain. This modification on the original user element allows us to simulate both the stable and unstable response of the magnetic beam during snapping, the results of which will be presented in Section~\ref{sec:beam_mag}. 

\subsection{Simulation procedure}
\label{sec:FEM_procedure}

We modeled an initially straight clamped-clamped beam as a 3D solid body. Similarly to the experiments (see Fig.~\ref{fig:Exp}), the beam was composed of two halves with antiparallel magnetization vectors. The length and width of the beam were the same as the experimental specimens.
The beam was discretized by the user-defined elements introduced in Section~\ref{sec:FEM_riks}, using a structured mesh with 16, 4, and 120 elements seeded, respectively, in the width, thickness, and length directions. The mesh was deemed to be sufficiently fine through a convergence study. The material of the beam was assumed incompressible with a shear modulus $G=0.39\,$MPa (paralleling the experiments; see Section~\ref{sec:experiments}) and a bulk modulus $100$ times larger than $G$. Given the large deflection of the beam during snapping, geometric nonlinearities were taken into account throughout the simulations. We highlight that the simulations employed the material properties characterized in the experiments (see Section~\ref{sec:experiments}), with no fitting parameters.

For the simulation protocol, we first imposed an end-to-end shortening, $\epsilon$, to buckle the beam and reach the preset bistable state. We then studied the snapping of the beam in three loading cases: (i) mechanical point load, (ii) pure magnetic load, and (iii) combined mechanical and magnetic loads. Each simulation run involved the following two sequential steps:

\emph{Step (a) -- Buckling:} First, $\epsilon$ was imposed to the straight beam, causing it to buckle into a curved configuration characterized by the classic sinusoidal Euler mode for a clamped-clamped beam. In this step, we obtained several post-buckled beam configurations by varying the end-to-end shortening $0\leq\epsilon\leq0.6$; the same range as in the experiments. To trigger buckling, geometric imperfections with the shape of the first eigenmode and a maximum amplitude of $0.1h$ were injected into the initial straight configuration.  

\emph{Step (b) -- Snapping:} \textit{For the loading case (i) -- (snap-through under mechanical point load)} we indented the beam at mid-span by prescribing the displacement, $\xi$, which was increased step-by-step until the beam reached the other stable configuration. The point load, $P$, was computed as a reaction force from equilibrium. From the load-displacement curve, $P(\xi)$, we identified the critical load for snapping at different end-to-end shortenings (Section~\ref{sec:beam_mech}). \textit{For loading case (ii) -- (snap-through under pure magnetic load)}, we applied a magnetic field on the entire beam with a \textit{Riks} step, in order to capture the full loading path during snapping. In the search for the equilibrium state, the strength of the applied field is set as an unknown, which, along with the displacements of the beam, was solved under a prescribed arc length increment of the loading path. We computed the critical field strength, the equilibrium path with stable and unstable branches, and the change of the strain energy during snapping. 
\textit{For the loading case (iii) -- (snap-through under combined mechanical and magnetic loads)}, a magnetic field with a given flux density lower than the critical value to trigger the snapping was first applied on the beam. Under this constant field, in the next step, we indented the beam by applying a displacement load at the mid-span to make it snap to the other stable configuration. We computed the indentation load-displacement curve, $P(\xi)$, for different values of the magnetic field. Then, we extracted the critical indentation force under the effect of magnetic load at different end-to-end shortenings.

\section{A reduced theoretical model for the snapping of magnetic beams}
\label{sec:theory}

We proceed by presenting a centerline-based theory for the  problem defined in Section~\ref{sec:problemdef} (see Fig.~\ref{fig:definition}b). We consider a thin, inextensible, hard-magnetic, and doubly-clamped beam, under Kirchhoff assumptions~\cite{o2017kirchhoff}; \textit{i.e.}, normals to the beam centerline remain normal and unstretched during deformation. Building upon recent developments for hard-magnetic beams~\cite{ciambella2020form, lum2016shape, wang2020hard, yan_comprehensive_2021,sano_kirchhoff-like_2022}, we develop a 1D beam model through dimensional reduction~\cite{audoly2010elasticity}, taking the 3D Helmholtz free energy for ideal hard-magnetic soft materials from Ref.~\cite{zhao2019mechanics} as a starting point, on top of other classic ingredients. The elastic (bending) energy of the beam is described by Euler's elastica~\cite{audoly2010elasticity}, and the work of indentation was addressed in Ref.~\cite{Pandey_2014_EPL}. Using the principle of virtual work (PVW), we will show that the derived ordinary differential equation (ODE) for the bending angle, $\theta(s)$, is equivalent to a clamped-clamped elastica under a {\it redefined} indentation load applied at mid-span ($s=L/2$). Hence, the  effect of the applied magnetic load on the snap-through buckling is qualitatively identical to that of a mechanical indentation force at mid-span. 

Following classic beam kinematics, we define $0\leq s\leq L$ to be the arc length of the (inextensible) centerline of a beam, located at $\mathbf{r}(s)=(x(s),y(s))$. We consider a  beam clamped both at $s=0$ and $s=L$, with $\mathbf{r}(0)=(0,0)$ and $\mathbf{r}(L)=(X,0)$. The bending angle, $\theta(s)$, is measured from $\hat{\mathrm{\mathbf{e}}}_x$, such that the centerline tangent is
$\hat{\mathbf{t}} \equiv \mathbf{r}' = (\cos\theta(s), \sin\theta(s)),$
where $(\cdot)' = \mathrm{d}(\cdot)/\mathrm{d}s$ and the corresponding boundary conditions are $\theta(0)=\theta(L)=0$. 
The relation between $\theta(s)$ and $\mathbf{r}(s)$ is obtained by integrating $\hat{\mathbf{t}}$:
\begin{eqnarray}
\mathbf{r}(L)=\left(x(L),\,y(L)\right) = \left(\int_0 ^L \cos\theta(s)\mathrm{d}s,\, \int_0 ^L \sin\theta(s)\mathrm{d}s\right) = \left( X,\, 0\right), \label{eq:const}
\end{eqnarray}
which acts as a constraint. 

Next, we consider first the external virtual work (EVW) and then the internal virtual work (IVW), before invoking the PVW to derive the governing equation for $\theta(s)$.

A reaction force $(F_x,\,F_y)$ is applied at $s = 0$ and the indentation load, $\mathrm{\mathbf{P}}=(0,\,-P)$, at $s=L/2$. Defining $\mathrm{\mathbf{N}}(s) = (N_x, N_y)$ as the internal force on the cross section at $s$, force balance yields
%
\begin{eqnarray}
\mathrm{\mathbf{N}}(s) = (N_x, N_y) = \left(- F_x,- F_y + P\Theta\left(s - \frac{L}{2}\right)\right),
\label{eq:Int_F}
\end{eqnarray}
with the Heaviside step function $\Theta(x)=\{{\rm sgn}(x)+1\}/2$ representing the discontinuity (of magnitude $P$) in the $N_{y}$ component at $s=L/2$, due to the applied indentation load. The EVW is then computed as
\begin{eqnarray}
\mathrm{EVW} &=& \int_0 ^L \left\{-F_x \cos\theta + \left(- F_y + P \Theta\left(s -\frac{L}{2}\right)\right)\sin\theta\right\}\mathrm{d}s.
\end{eqnarray}
%

The Helmholtz free energy proposed in Ref.~\cite{zhao2019mechanics} for hard-magnetic materials can be decomposed into an elastic part, associated with mechanical deformation, and a magnetic part, arising from the interactions between remanent magnetization and the external field. Based on this decomposition, the total energy of a hard-magnetic beam is the sum of the elastic energy, $U^\mathrm{el}$, and the magnetic potential, $U^\mathrm{m}$. Assuming a Hookean constitutive law,
\begin{eqnarray}
U^{\rm el} = \int_0 ^L \frac{EI}{2}\theta'^{2}\mathrm{d}s,
\label{Benergy}
\end{eqnarray}
where $EI$ is the bending stiffness of the beam of Young's modulus, $E$, and an area moment of inertia, $I = h^3b/12$. 

According to~\cite{zhao2019mechanics}, we now make use of the magnetic potential density $\hat{U}^\mathrm{m}$ in Eq.~(\ref{Um_3D_beam}).
Focusing on the geometry of our problem (see Fig.~\ref{fig:definition}b), we set the applied field to $\mathrm{\mathbf{B}}^{\mathrm{a}} = B^{\mathrm{a}}\hat{\mathrm{\mathbf{e}}}_y$, and the magnetization vector  $\mathrm{\mathbf{M}}=\mathrm{\mathbf{B}}^\mathrm{r}/\mu_0$ exhibiting the specific profile of Eq.~(\ref{eq:magnetization}); in the deformed configuration, $\mathrm{\mathbf{M}}$ is parallel to the tangent vector $\hat{\mathrm{\mathbf{t}}}$ for $0 \leq s \leq L/2$ (or anti-parallel for $L/2 \leq s \leq L$). \textcolor{black}{The deformation gradient ${\mathbf{\stackrel{\circ}{F}}}$ for thin beams has been derived in~\cite{lum2016shape,wang2020hard,yan_comprehensive_2021} as:}

\begin{equation}
{\mathbf{\stackrel{\circ}{F}}}=\left(\begin{array}{cc}
\cos \theta & -\sin \theta \\
\sin \theta & \cos \theta
\end{array}\right).
\end{equation}
Hence, the magnetic potential for our beam is
\begin{eqnarray}
U^{\mathrm{m}} &=& -\int_0 ^L hb \mathbf{\stackrel{\circ}{F}}\mathrm{\mathbf{M}}\cdot\mathrm{\mathbf{B}}^{\mathrm{a}} \mathrm{d}s = \frac{hbB^{\rm r}B^{\rm a}}{\mu_0}\int_0 ^L {\rm sgn}\left(s - \frac{L}{2}\right)\sin\theta(s) \mathrm{d} s.
\end{eqnarray}

Invoking the PVW, mechanical equilibrium is assured when the EVW is balanced by the IVW $ = \delta U^{\rm el} + \delta U^{\rm m}$ yielding 
\begin{eqnarray}
EI\theta'' + F_x\sin\theta - \tilde{F}_y\cos\theta = -\frac{1}{2}\left(P-\frac{2hbB^{\rm a}B^{\rm r}}{\mu_0}\right){\rm sgn}\left(s - \frac{L}{2}\right)\cos\theta\label{eq:ode_theta},
\label{ode}
\end{eqnarray}
with $\tilde{F}_y\equiv F_y-(P/2)$ and boundary conditions $\theta(0) = \theta(L)= 0$. The two unknowns, $F_x$ and $F_y$, are the Lagrange multipliers associated with the clamped boundary~\cite{audoly2010elasticity} and can be determined through Eq.~(\ref{eq:const}). Note that the term in Eq.~(\ref{eq:ode_theta}) involving ${2hbB^{\rm a}B^{\rm r}}/{\mu_0}$ can be interpreted an a second indentation force, in addition to $P$. Therefore, we can define an effective indentation force under the combined mechanical and magnetic loading:
\begin{eqnarray} \label{eq:pstar}
P^{*} \equiv P-\frac{2hbB^{\rm a}B^{\rm r}}{\mu_0},
\end{eqnarray}
used to rewrite Eq.~(\ref{eq:ode_theta}) as
\begin{equation}\label{eq:ode_theta2}
EI\theta'' + F_x\sin\theta - \tilde{F}_y\cos\theta = -\frac{P^{*}}{2}{\rm sgn}\left(s - \frac{L}{2}\right)\cos\theta.
\end{equation}
This new ODE is equivalent to a clamped-clamped elastica under indentation force $P^{*}$ applied at $s=L/2$~\cite{vangbo_lateral_1998, Pandey_2014_EPL}. Under appropriate boundary conditions, $\theta(0)=\theta(L)=0$, and the constraint in Eq.~(\ref{eq:const}), Eq.~(\ref{eq:ode_theta2}) defines a boundary value problem that can be solved numerically to predict the classic N-shape snap-through response of a doubly clamped beam~\cite{vangbo_lateral_1998}, but now under combined magnetic and indentation loading. We do so using the solver \texttt{bvp5c} in~\texttt{MATLAB}. Note that all the relevant parameters in this model are characterized experimentally, and there are no fitting parameters. In Sections~\ref{sec:beam_mech}--\ref{sec:Beam_mechmag}, we will compare the predictions from this magnetic beam model against FEM (developed in Section~\ref{sec:numerics}) and experiments (developed in Section~\ref{sec:experiments}).

\subsection{Linearized theory with $\epsilon \ll 1$ and $|\theta| \ll 1$}
\label{sec:linear_theory}

For configurations of the bistable beam with small values of end-to-end shortening ($\epsilon \ll 1$), the deformations are small ($|\theta| \ll 1$) at the onset of snapping. In this limit, with $\sin \theta \simeq \theta$ and $\cos \theta\simeq1-({\theta^2}/{2})$, Eq.~(\ref{eq:ode_theta2}) simplifies to
\begin{equation} \label{simpODE}
\theta''+\overline{F}_x\theta-\overline{\tilde{F}}_y=-\frac{\overline{P}^{*}}{2}{\rm sgn}\left(\overline{s}-\frac{1}{2}\right),
\end{equation}
where we have used the following dimensionless variables: $\overline{s}=s/L$, $\overline{F}_x\equiv{F_x L^2}/{EI}$, $\overline{\tilde{F}}_y\equiv{\tilde{F}_y L^2}/{EI}$, and $\overline{P}^{*} \equiv {PL^2}/{EI}-({2 h b B^{\mathrm{a}}/ B^{\mathrm{r}} L^2}{EI \mu_{0}})$. Expanding Eq.~(\ref{eq:const}) with respect to $|\theta|\ll1$, the corresponding boundary conditions are
\begin{equation}
\int_{0}^{1} \frac{\theta^{2}(\overline{s})}{2} \mathrm{d} \overline{s}=\epsilon, \quad\mathrm{and}\quad \int_{0}^{1} \theta(\overline{s}) \mathrm{d}\overline{s}=0.      
\label{eq:bc}
\end{equation}
Employing the method of variation of parameters, the solution of Eq.~(\ref{simpODE}) is
\begin{equation}
   \theta(\bar{s})=\frac{\overline{\tilde{F}}_{y}}{\kappa^{2}} \varphi_{1}(\overline{s})+\frac{\overline{P}^{*}}{\kappa^{2}} \varphi_{2}(\overline{s}) ,
   \label{eq:lin_sol}
\end{equation}
where we have introduced the wavenumber $\kappa \equiv \sqrt{\overline{F}_{x}}$ and the functions $\varphi_1$ and $\varphi_2$, which are respectively, symmetric and asymmetric functions with respect to ${s}=1/2$, are defined in~\ref{sec:Var Parameters}. 
The two unknown parameters, $\kappa=\sqrt{\overline{F}_{x}}$ and $\overline{\tilde{F}}_y$, are determined using the boundary conditions in Eq.~(\ref{eq:bc}) to arrive at
\begin{equation}
\kappa=9, \quad\mathrm{and}\quad   2\epsilon=\left(\frac{\overline{\tilde{F}}_{y}}{\kappa^{2}}\right)^{2} c_{1}+\left(\frac{\overline{P}^{*}}{\kappa^{2}}\right)^{2} c_{2},   
\label{ode}
\end{equation}
where $c_1\equiv\int_0 ^1(\varphi_1(\overline{s})) ^2\mathrm{d}\overline{s}$ and $c_2\equiv\int_0 ^1(\varphi_2(\overline{s})) ^2\mathrm{d}\overline{s}$ are two positive numerical constants detailed in~\ref{sec:Var Parameters}. 

Using Eqs.~(\ref{ode}), we can now discuss the critical condition for snap buckling. Given that $c_1$, $c_2$ and $\epsilon$ are all positive, $\overline{\tilde{F}}_y$ ceases to exist, and the beam snaps, when 
$(\overline{P}^{*}/\kappa^{2})^{2} c_{2} \geq 2 \epsilon$.
Hence, the critical condition for the snap-transition is 
\begin{equation}
|\overline{P}^{*}_\mathrm{cr}|=\kappa^{2} \sqrt{\frac{2}{c_{2}} \epsilon}=C_0 \sqrt{\epsilon},
\label{Eqsnap}
\end{equation}
with the positive constant $C_0 \simeq 130$~\cite{Pandey_2014_EPL,vangbo_lateral_1998}. 
The dimensional version of Eq.~(\ref{Eqsnap}), through Eq.~(\ref{eq:pstar}), is
\begin{equation} \label{eq:snap_dimensional}
\left|\frac{bh B^{\mathrm{a}} B^{\mathrm{r}} L^{2}}{\mu_{0} EI}-\frac{P_\mathrm{cr} L^{2}}{2 EI}\right|=C_0 \sqrt{\epsilon}.
\end{equation}
For reasons that will become clearer in Section~\ref{sec:Beam_mechmag}, we recognize the critical load at which snapping occurs as the maximum load that the beam can support before snap-through, $P_\mathrm{cr}=P_\mathrm{max}$, and we rewrite Eq.~(\ref{eq:snap_dimensional}) back in dimensionless form
\begin{equation} \label{eq:odeSolve}
 \overline{P}_\mathrm{max}=2 (\overline{B}^\mathrm{a} + C_0 \sqrt{\epsilon}),
\end{equation}
where $\overline{P}_\mathrm{max}=P_\mathrm{max}L^2/(EI)$ and the applied magnetic field was nondimensionalized as:
\begin{equation} \label{eq:nonD-B}
  \overline{B}^a=\frac{bhL^2B^\mathrm{a}_\mathrm{cr} B^\mathrm{r}}{EI\mu_\mathrm{0}},
\end{equation}
characterizing the relative importance between the magnetic load and beam-bending effects. According to Eq.~(\ref{eq:odeSolve}), under the assumption of $\epsilon\ll 1$ and when $\overline{B}^a>0$ (\textit{i.e.}, $\mathbf{B}^\mathrm{a}$  and $\mathbf{P}$ are in the same direction), the maximum indentation force that the clamped-clamped magnetic beam can support before snapping is expected to depend linearly on the applied magnetic field $\overline{B}^a$, with a slope 2 and an offset $2C_0\sqrt{\epsilon}$ set by the end-to-end shortening. This prediction will be tested against experiments and FEM in Fig.~\ref{fig:Fmax_B} (Section~\ref{sec:Beam_mechmag}). In the absence of a magnetic field ($\overline{B}^a=0$), we recover the standard result, the critical indentation force for purely elastic snapping~\cite{Pandey_2014_EPL}, with the scaling $\overline{P}_\mathrm{max}\sim\sqrt{\epsilon}$, which will be tested against experiments and FEM in Fig.~\ref{Fig:Mech_load} (in Section~\ref{sec:beam_mech}).

\section{Snapping under mechanical point loading}
\label{sec:beam_mech}

\begin{figure}[h!]
    \centering
    \includegraphics[width=\columnwidth]{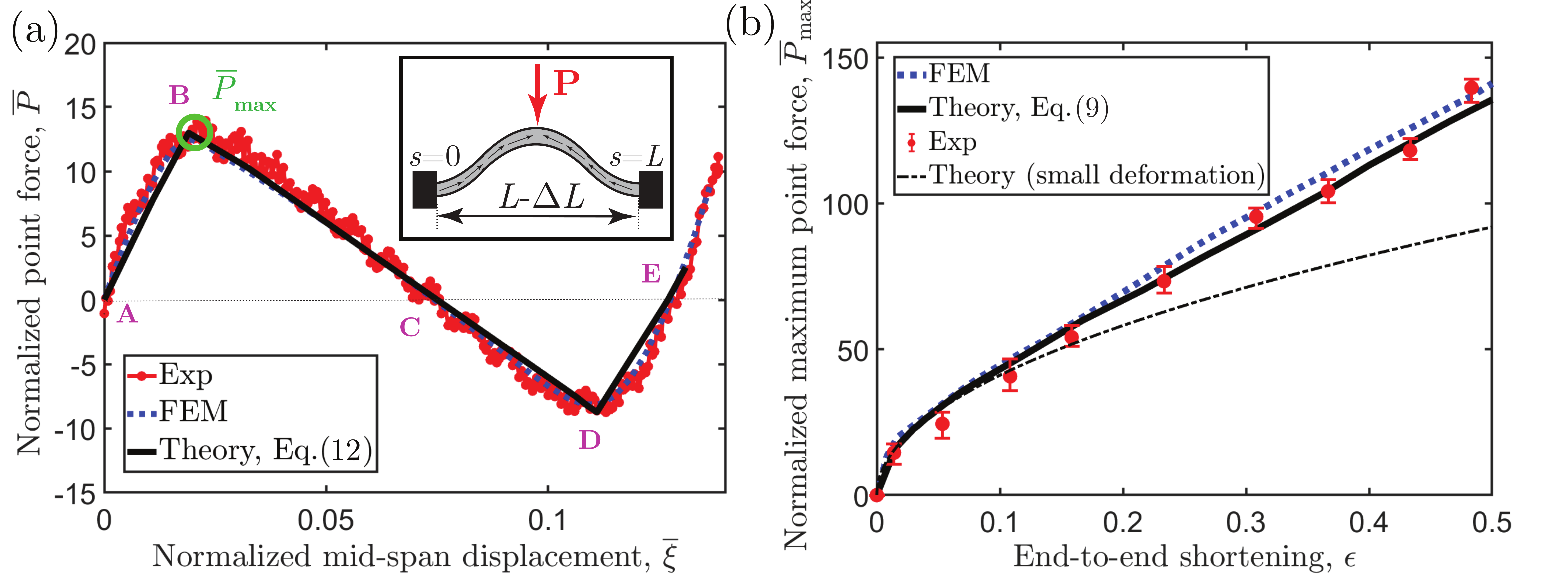}
    \caption{Snap buckling under mechanical point load ($B^a=0$). (a) Normalized mechanical point force, $\overline{P}$, versus the mid-span displacement, $\overline{\xi}$, for the end-to-end shortening of $\epsilon=0.014$. The maximum of the curve is defined as $\overline{P}_\mathrm{max}$. (b) Normalized maximum point force, $\overline{P}_\mathrm{max}$, versus $\epsilon$. The error bars correspond to the standard deviations of the experimental measurements for three identical specimens. (Inset) Schematic diagram of the loading configuration. 
    }
    \label{Fig:Mech_load}
\end{figure}

We start by focusing on the classic bistable response when the beam subjected \textit{only} to a mechanical point load ($B^a=0$) at mid-span. Even if well-established~\cite{Pandey_2014_EPL}, this case serves as a pre-validation of the framework against experiment, before introducing magnetic effects in Section~\ref{sec:beam_mag}. In Fig.~\ref{Fig:Mech_load}a, we present the results for the dimensionless indentation force, $\overline{P}=PL^2/EI$, versus the dimensionless mid-span displacement, $\overline{\xi}=\xi/L$. The initial buckled configuration was generated with an end-to-end shortening of $\epsilon=0.014$.
Then, $\overline{\xi}$ was gradually increased while measuring the indentation force. The resulting $\overline{P}(\overline{\xi})$ force-displacement curve exhibits the classic N-shape representative of bistable mechanisms~\cite{vangbo_lateral_1998}. Points A and E are the two stable stages. The maximum normalized indentation force, $\overline{P}_\mathrm{max}$, occurs at point B. The unstable branch, with  negative stiffness, occurs between points B and D, and plot C is the unstable equilibrium state. Excellent agreement is found between experiments, FEM, and the solution of Eq.~(\ref{eq:ode_theta2}).

In Fig.~\ref{Fig:Mech_load}b, we plot $\overline{P}_\mathrm{max}$, as a function of $\epsilon$, finding a sub-linear dependence.
For small values of the end-to-end shortening ($\epsilon\lesssim0.1$), the observed scaling $\overline{P}_\mathrm{max}\sim\sqrt{\epsilon}$ (dot-dashed line in Fig.~\ref{Fig:Mech_load}c) is consistent with Eq.~(\ref{eq:odeSolve}) when $B^a=0$, obtained from the linearized theory for small deformations. For $\epsilon \gtrsim 0.1$, the linearized theory no longer works, but the nonlinear theory of Eq.~(\ref{eq:ode_theta2}) with $B^{\rm a}=0$ (solid line in Fig.~\ref{Fig:Mech_load}c) is in excellent agreement with the FEM and experiments, through the full range of explored $\epsilon$. This agreement between the experiments, FEM, and the reduced-order beam model, even if within a classic setting, serves as a first step in validation.

\section{Snapping under magnetic loading}
\label{sec:beam_mag}

We proceed by investigating the buckling of the bistable beam under an external magnetic field, this time with no mechanical load ($P=0$), seeking to quantify how the  critical  magnetic field strength, $B^\mathrm{a}_\mathrm{cr}$, required for switching between the two stable states, depends on the end-to-end shortening, $\epsilon$.

\begin{figure}[h!]
    \centering
    \includegraphics[width=0.6\columnwidth]{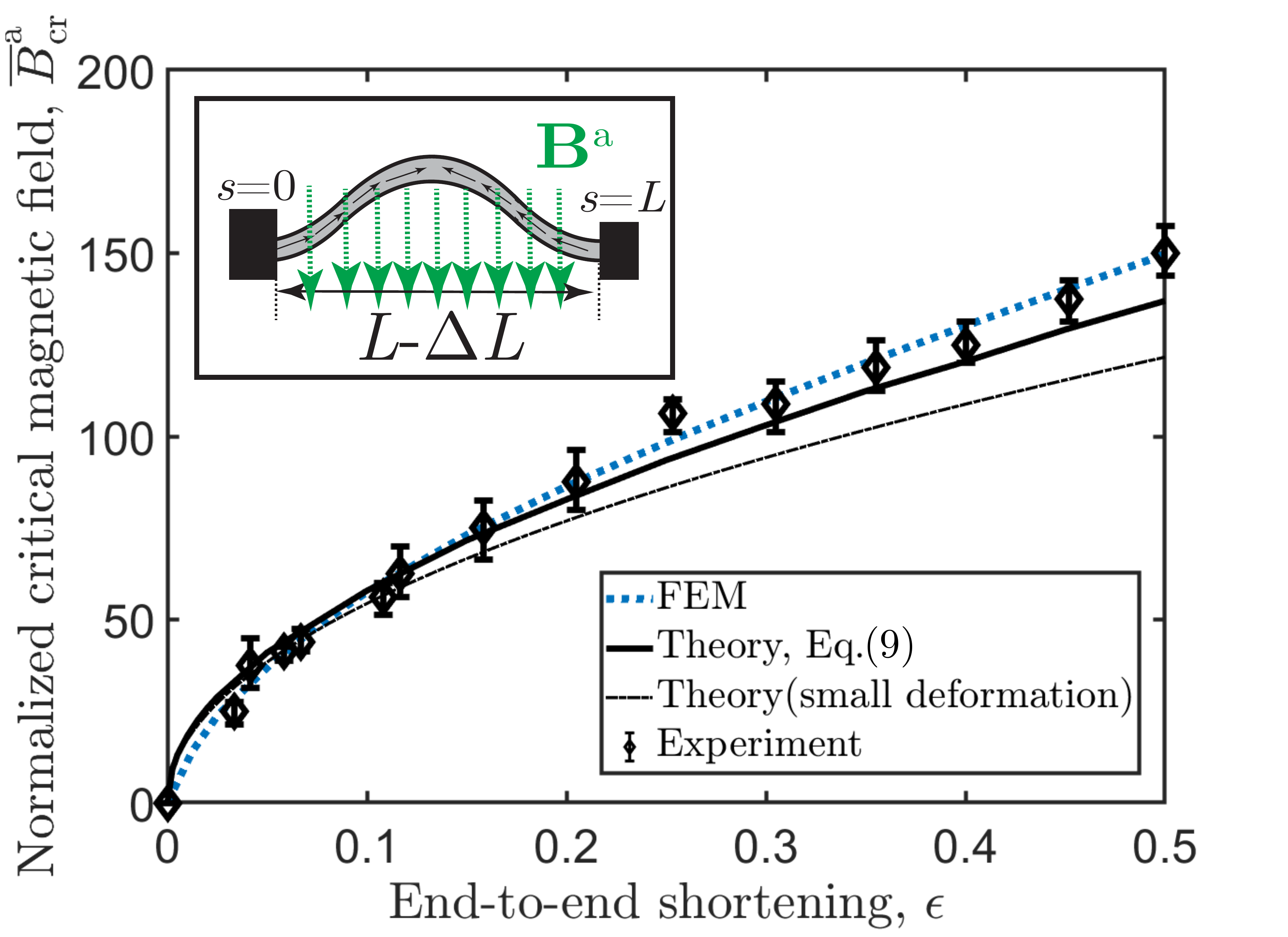}
    \caption{Snap buckling under magnetic actuation. Normalized critical strength of the uniform magnetic field required for beam snapping, $\overline{B}^\mathrm{a}_\mathrm{cr}$, as a function of end-to-end shortening, $\epsilon$. The results were obtained from the nonlinear elastic theory in Eq.~(\ref{eq:ode_theta2}) (solid line), small-deformations theory (dashed line), FEM (dotted line), and experiments (data points with error bars). The error bars of the experimental data correspond to the standard deviation of the measurements on three identical specimens. (Inset) Schematic of the bistable beam under magnetic load loading.}
    \label{fig:critical_B}
\end{figure}

In Fig.~\ref{fig:critical_B}, making use of the dimensionless magneto-elastic parameter defined in Eq.~(\ref{eq:nonD-B}), we plot $\overline{B}^\mathrm{a}_\mathrm{cr}(\epsilon)$ curves obtained as predicted from FEM simulations, the 1D theory and the experiments. Naturally, increasingly deformed pre-configurations (increasing $\epsilon$) requires a higher value of $B^\mathrm{a}_\mathrm{cr}$ for snapping. We find a good agreement between the FEM, the experimental data, and the solution of Eq.~(\ref{eq:ode_theta2}). For small deformations ($\epsilon\lesssim0.1$), the data follows the scaling $B^\mathrm{a}_\mathrm{cr} \sim \sqrt{\epsilon}$, consistently with Eq.~(\ref{eq:odeSolve}). For higher values of $\epsilon$, the overall $B^\mathrm{a}_\mathrm{cr}(\epsilon)$ curves computed from FEM, are captured by the solutions of Eq.~(\ref{eq:ode_theta2}) with $P=0$ reasonably well. 


\begin{figure}[h!]
    \centering
    \includegraphics[width=\columnwidth]{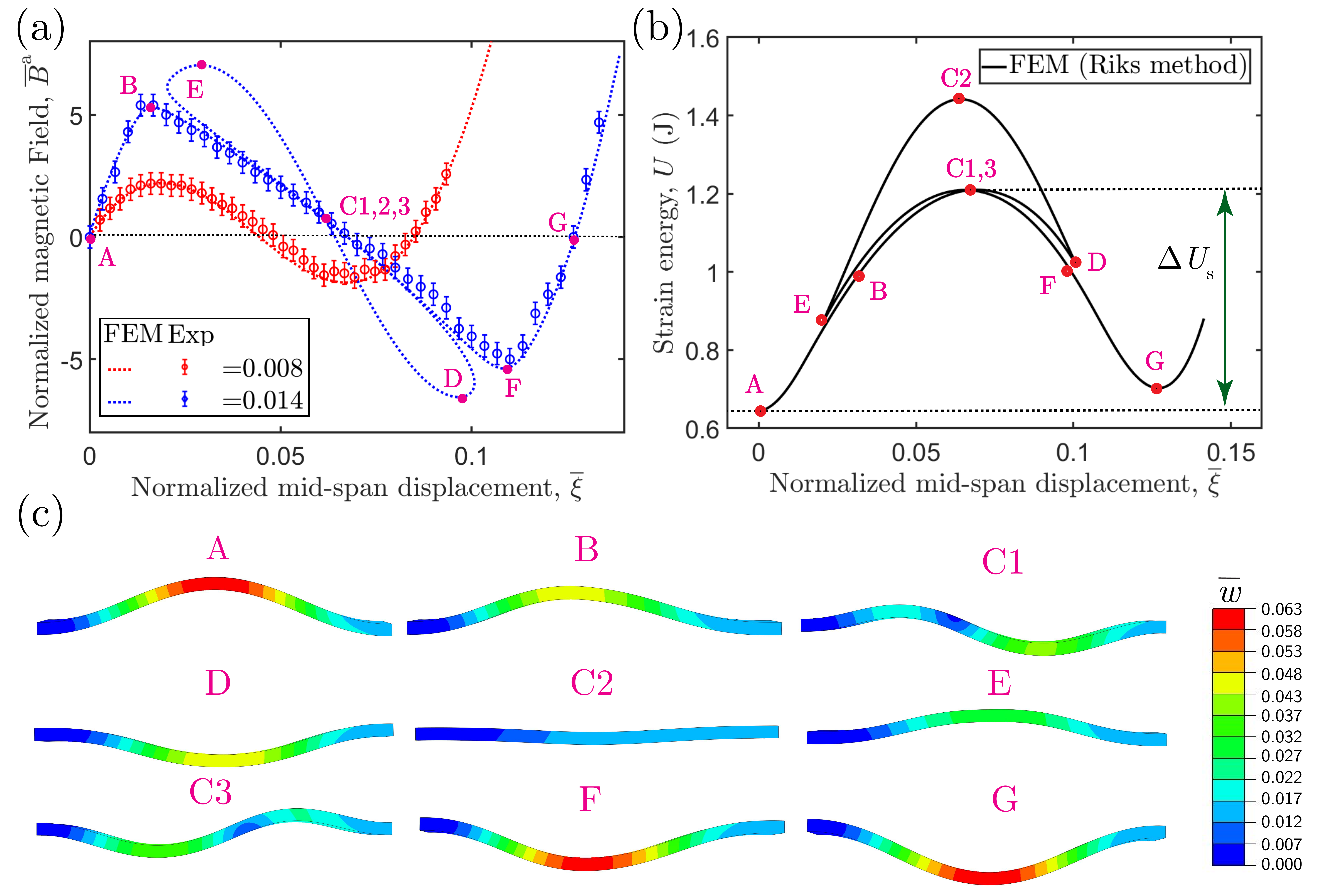}
    \caption{Bifurcation diagram for the snap-through of the bistable beam under magnetic actuation. (a) Normalized magnetic field, $\overline{B}^a$, versus the normalized mid-span displacement, $\overline{\xi}$: FEM simulations and experiments for two beams with $\epsilon=\{0.008,\,0.014\}$. (b) FEM-computed strain energy, $U$, versus $\overline{\xi}$, for $\epsilon=0.014$. The beam must overcome the energy barrier, $\Delta U_s$ (vertical double-arrow), to switch between the two stable states. (c) Representative FEM-computed configurations of the beam along the equilibrium solution path, corresponding to the same points in the plots of panels (a) and (b). The colorbar represents the normalized displacement of the beam $\overline{w}$.}
    \label{fig:riks}
\end{figure}

The Riks procedure in FEM (cf. Section~\ref{sec:FEM_procedure}), enables us to capture the unstable equilibrium path during snapping under actuation by a magnetic field, between the first to the second stable configuration. 
In Fig.~\ref{fig:riks}a, we plot the normalized magnetic load, $\overline{B}^\mathrm{a}$, as a function of the normalized mid-span displacement of the beam, $\overline{\xi}$, for two representative values of the end-to-end shortening, $\epsilon=\{0.008,\,0.014\}$. The 
FEM-computed results (dotted lines) are in quantitative agreement with the experimental data. For $\epsilon=0.008$, the $\overline{B}^\mathrm{a}(\overline{\xi})$ curve is non-monotonic, first increasing to a maximum, then decreasing to become negative, until a minimum is reached, to then increase again. The case with $\epsilon=0.014$ is more complex; the Riks method captures a force‐displacement equilibrium path with a complex transition between the two stable states, with winding branches and  multiple equilibrium solutions for a the same $\overline{\xi}$. However, note that some of the winding-branch segments computed from FEM are not practically relevant; only the solutions with the lowest energy barrier are experimentally observable.

To gain further insight into the energetics of the load-displacement path discussed above, focusing on $\epsilon=0.014$, we now use FEM to compute the total strain energy, $U$, as a function of $\overline{\xi}$ during the snapping process; the results are plotted in Fig.~\ref{fig:riks}b. The points A, B,\ldots, G labeled in the plot correspond to the computed configurations shown in Fig.~\ref{fig:riks}c. During the transition path between the stable states A and G, $U$ increases with $\overline{\xi}$ from a minimum (A) to a local maximum (C1) and decreases to another minimum (D). Hence, the corresponding energy barrier, $\Delta U_\mathrm{s}$, between this minimum and the local maximum must be overcome for snap-through. 
According to the principle of minimum potential energy, the lowest-energy path is the one observed in practice. Consequently, the higher energy configurations shown in Fig.~\ref{fig:riks}c for points D, C2, E, and C3 are not observed experimentally. Indeed, the experimentally observed path in Fig.~\ref{fig:riks}a is an excellent match with the latest-energy path of Fig.~\ref{fig:riks}b, passing through the points A-B-C1-F-G. 


\section{Snapping under combined mechanical and magnetic loading}
\label{sec:Beam_mechmag}

Finally, we turn to the combined case of simultaneously loading the bistable beams with a mechanical indentation and a magnetic field, each of each was tackled individually in the previous Sections~\ref{sec:beam_mech} and~\ref{sec:beam_mag}. We seek to quantify how the magnetic loading modifies the load-bearing capacity of the bistable beam under indentation and characterize the critical conditions for snap buckling.

\begin{figure}[h!]
    \centering
    \includegraphics[width=0.9\columnwidth]{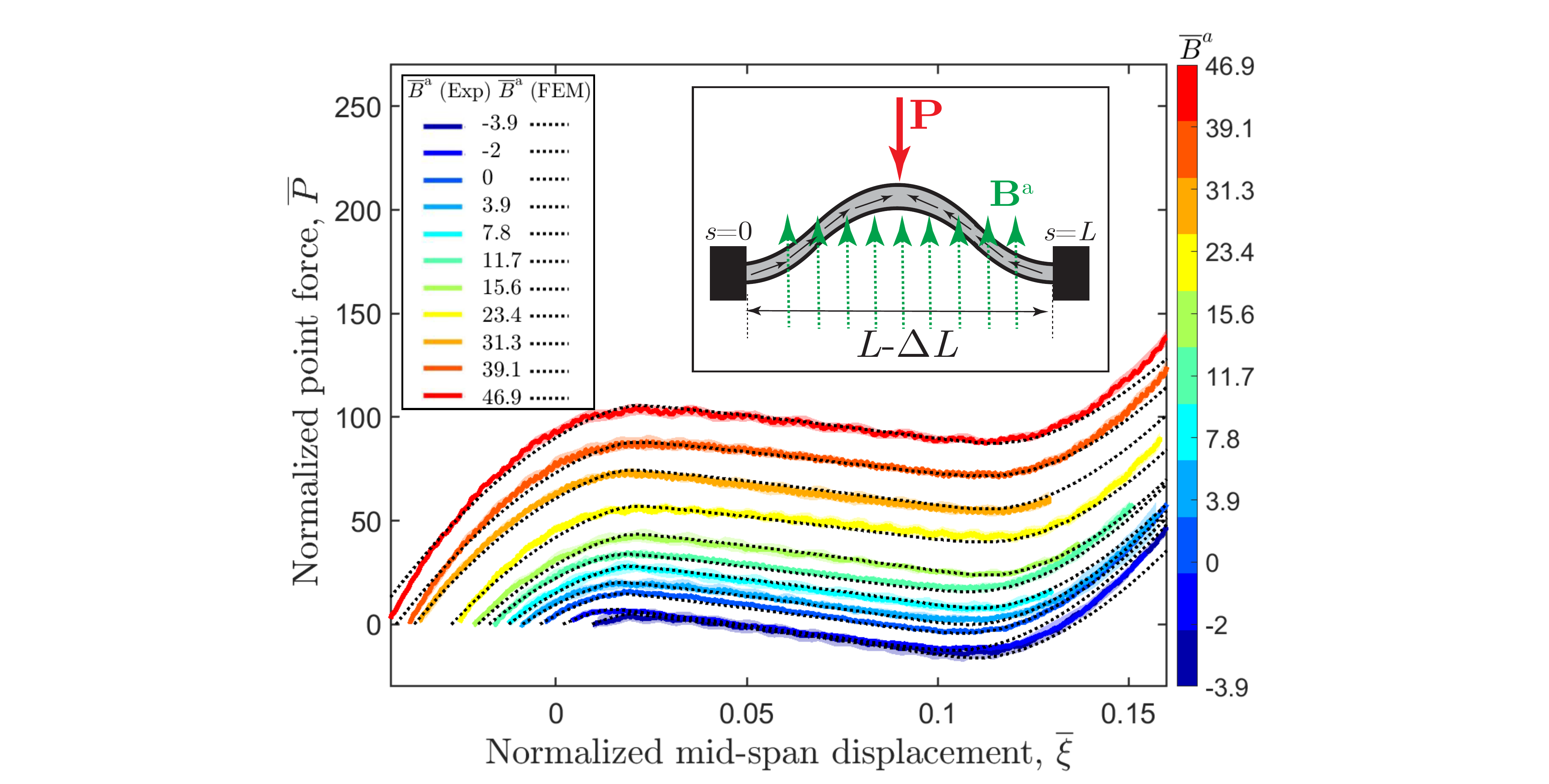}
    \caption{Force–displacement curves for the  indented magnetic beam (with $\epsilon=0.014$) in the presence of a uniform magnetic field. The normalized indentation force, $\overline{P}$, is plotted versus the normalized mid-span displacement, $\overline{\xi}$, at various levels of prescribed field strength, $\overline{B}^\mathrm{a}$ (see legend): experiments (solid lines) and FEM simulations (dotted lines). The shaded region of each curve represents the standard deviation of three identical measurements. (Inset) Schematic diagram of the bistable beam under combined magnetic and mechanical loading.}
    \label{fig:FW_variousB}
\end{figure}

In Fig.~\ref{fig:FW_variousB}, we present the normalized indentation force, $\overline{P}$, versus the beam's mid-span displacement, $\overline{\xi}$, at different levels of the uniform magnetic field, $\overline{B}^a \mathbf{\hat{e}}_y$, varied systematically in the range $-3.9 \leq \overline{{B}}^\mathrm{a} \leq 46.9$ (see legend of the plot). We focus on the representative case with $\epsilon=0.014$. To track the full equilibrium path, including its unstable portions, the indenter was glued to the beam at mid-span, as described in Section~\ref{sec:procol}.  At each value of $\overline{{B}}^\mathrm{a}$, the signal-to-noise ratio of the measurements was enhanced by repeating three independent, but otherwise identical, experimental runs; their average is reported as the $\overline{P}(\overline{\xi})$ curves of Fig.~\ref{fig:FW_variousB}. Throughout, excellent agreement is found between experiments (solid lines) and the FEM (dotted lines). 

Under an external magnetic field, the force-indentation response of the hard-magnetic beam can be modified significantly with respect to the purely mechanical case ($\overline{B}^\mathrm{a}=0$ and results in Section~\ref{sec:beam_mech}). When the applied magnetic field is in the opposite direction of the indentation force ($\overline{B}^\mathrm{a}>0$), the generated magnetic torques oppose the direction of the indentation-induced beam rotation. Consequently, as the field strength is increased, the beam stiffens and becomes more resistant to snap buckling (the local maximum of $\overline{P}$ increases). By contrast, when the magnetic load is applied in the same direction to the indentation ($\overline{B}^\mathrm{a}<0$), snap-through occurs at lower indentation forces as the magnetic torques are in the same direction of the indentation-induced rotation.

\begin{figure}[h!]
    \centering
    \includegraphics[width=\columnwidth]{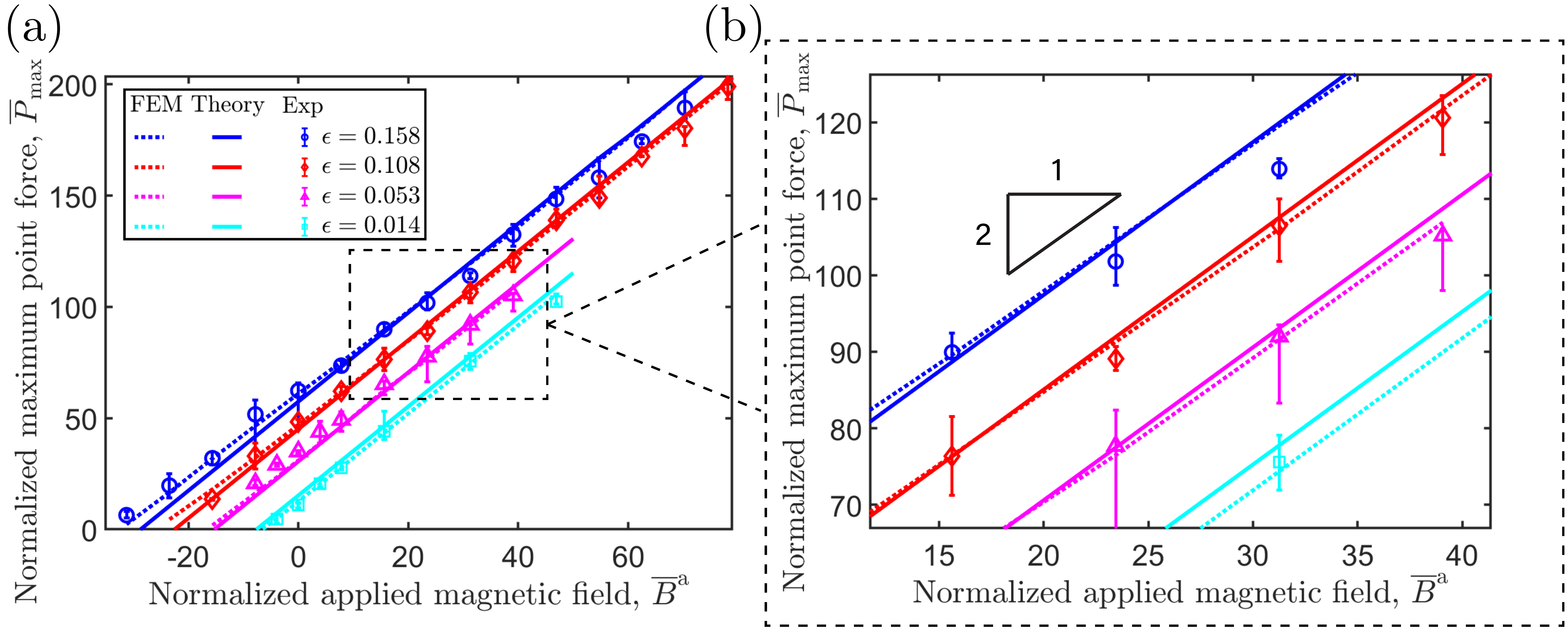}
    \caption{Critical indentation load for snap-transition under magnetic actuation. (a) Normalized maximum indentation force, $\overline{P}_\mathrm{max}$, versus the prescribed applied magnetic field, $\overline{B}^a$, at different end-to-end shortenings ($\epsilon=\{0.158, 0.108, 0.053, 0.014\}$): experiments (data points), theoretical predictions from Eq.~(\ref{eq:ode_theta2}) (solid lines), and FEM (dotted lines). (b) Zoomed view of (a) with a depicted slope of 2, consistently with Eq.~(\ref{eq:odeSolve}).}
    \label{fig:Fmax_B}
\end{figure}

In Fig.~\ref{fig:Fmax_B}a, we quantify the dimensionless relation between the maximum (critical) indentation force, $\overline{P}_\mathrm{max}$, and the magnitude of the prescribed magnetic field $\overline{B}^\mathrm{a}$, for four values of the end-to-end shortenings, $\epsilon=\{0.014,$ $0.053,$ $0.108,$ $0.158\}$. The experimental force-displacement signals were smoothed with a 50-point moving average filter, to facilitate the  extraction of $\overline{P}_\mathrm{max}$, the largest load that the beam can sustain prior to snapping. Again, an excellent agreement is observed between the experiments (data points), FEM (dotted lines), and the solution of Eq.~(\ref{eq:ode_theta2}) (solid lines). We find the robust linear scaling  $\overline{P}_\mathrm{max}\sim\overline{B}^\mathrm{a}$, with a slope of 2 (see zoomed inset in Fig.~\ref{fig:Fmax_B}b). Increasing $\epsilon$ results in an increase of the offset at $\overline{B}^\mathrm{a}=0$ of the linear curves, as dictated by the purely mechanical indentation case in Fig.~\ref{Fig:Mech_load}. Hence, increasing the end-to-end shortening results in a larger indentation force required for snapping under a particular level of field strength, set by the offset of the linear behavior. The experimental and FEM data are in remarkable agreement with Eq.~(\ref{eq:odeSolve}), indicating that the largest value of end-to-end shortening explored in these experiments (and FEM simulations) still lies in the regime of validity, with small deformations ($|\theta| \ll 1$), of the linearized theory developed in Section~\ref{sec:linear_theory}. 

\begin{figure}[h!]
    \centering
    \includegraphics[width=0.6\columnwidth]{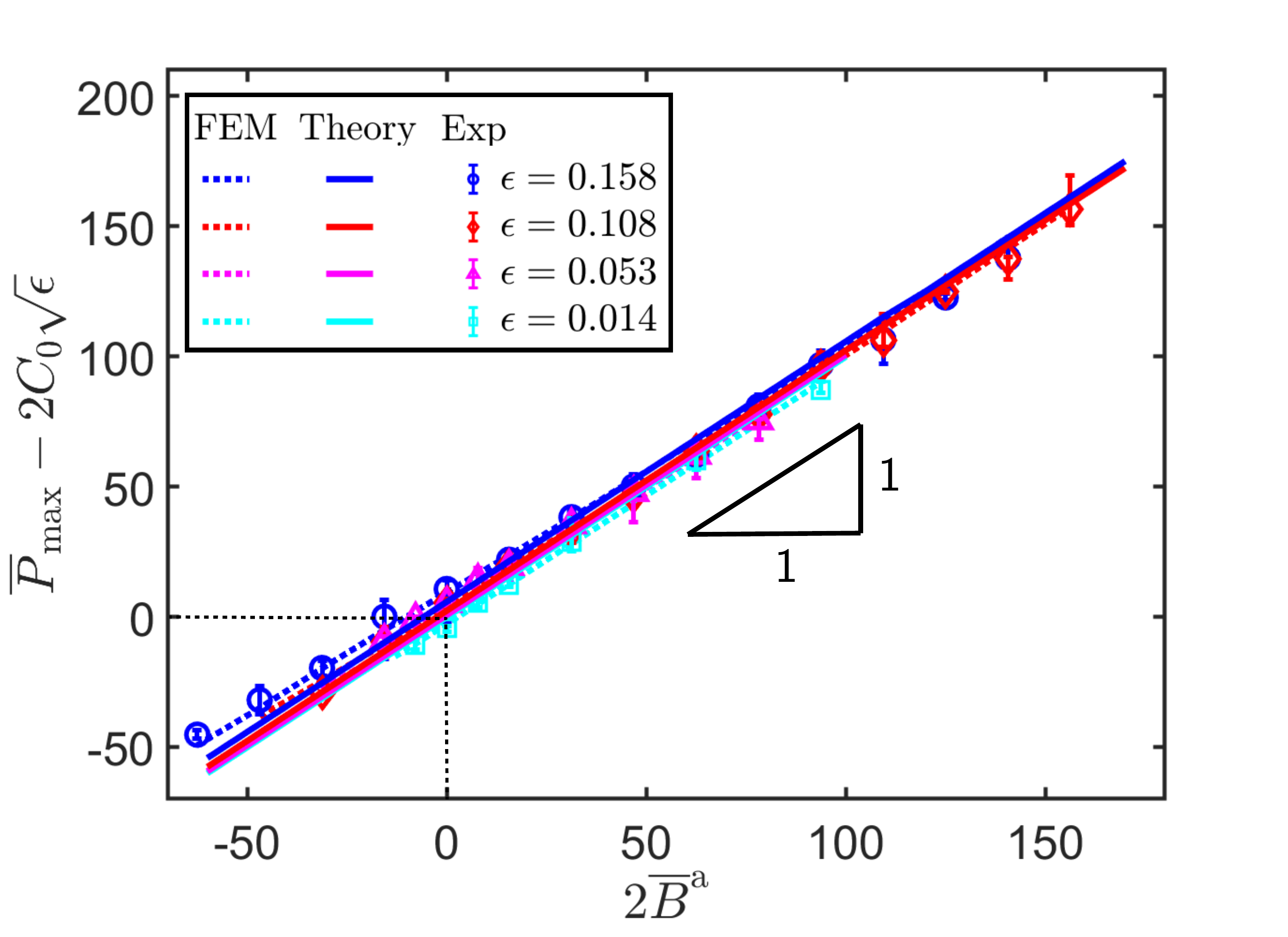}
    \caption{Master curve for the experimental results, FEM simulations, and the theoretical predictions, for the data in Fig.~\ref{fig:Fmax_B}. The quantity, $\overline{P}_\mathrm{max}-2C_0 \sqrt{\epsilon}$ in Eq.~(\ref{eq:odeSolve}), is plotted as a function of $2 \overline{B}^\mathrm{a}$, yielding a collapse of all the data. The term $-2C_0\sqrt{\epsilon}$ removes the offset due to the end-to-end shortening.} 
    \label{fig:PB_collapse}
\end{figure}

In Fig.~\ref{fig:PB_collapse}, making use of Eq.~(\ref{eq:odeSolve}), we now replot all the data in Fig.~\ref{fig:Fmax_B} but with $\overline{P}_\mathrm{max}-2C_0 \sqrt{\epsilon}$ as a function of $2 \overline{B}^\mathrm{a}$. The purpose is to remove the effect of end-to-end shortening, and characterize the magneto-elastic effect of the snapping beam. As predicted by the linearized theory, we find a striking collapse of all the data into a master curve of unit slope, passing through the origin. This collapse indicate that Eq.~(\ref{eq:odeSolve}), based on a linearized theory and combining the dimensionless groups of magnetic (${bh B^\mathrm{a} B^\mathrm{r} L^2}/{\mu_0 EI}$) and mechanical (${PL^2}/{EI}$) load, serves as a high-fidelity description of the magneto-elastic behavior of our hard-magnetic bistable beams, with different end-to-end shortenings ($\epsilon\lesssim 0.1$), in the limit of small deformations ($\theta \ll 1$).

\section{Conclusion}
\label{sec:conclusion}
We investigated the snapping behavior of bistable magneto-active beams under combined mechanical and magnetic actuation, incorporating experiments, FEM, and a reduced-order model. Considering a pre-compressed bistable beam with different levels of end-to-end shortening, we characterized the load-displacement response, the critical indentation force, the field strength at the onset of snapping, and the effect of magnetic loading on snap buckling under indentation. The Riks method was employed in the 3D FEM simulations to analyze the snap transition. We also developed a beam theory to rationalize the observed magneto-elastic response. Precision experiments validated the theory and the FEM simulations. 

More specifically, we studied the snap buckling of the beam under three different loading cases: (i) mechanical point load only, (ii) magnetic field only, and (iii) combined magnetic and mechanical loading. Case (i), even if classic, served for pre-validation. In case (ii), we triggered snap buckling under a magnetic field for various end-to-end shortenings, by designing the magnetization profile of the beam. For small-deformations, the critical magnetic field increased with the square root of the end-to-end shortening. The Riks method was used to explore the equilibrium transition path, finding that increasing the end-to-end shortening complicates the instability response, but the experimentally observable solution corresponds to the lowest energy level. Finally, in case (iii), we examined how magnetic loading affects the indentation-induced snapping of the bistable beam. The critical indentation force for snapping can be adjusted by the magnitude and direction of the magnetic field. Our magnetic beam model captures these results. In the small deformation limit, the critical indentation force at the onset of snapping is linearly proportional to the applied magnetic field with a slope and offset that can be predicted. In this limit, a master curve is uncovered that collapses the experimental and FEM-computed data.

Our study provides insight into the nonlinear magneto-elastic coupling of bistable beams, which could be extended in several directions for future work. Fundamentally, the dynamics of multi-stable structures integrated with soft active materials remain relatively unexplored and deserve further attention. From a practical viewpoint, optimization and inverse design is an exciting direction: compact actuators could be designed using bistable beams while minimizing the total energy consumption during actuation. The magnetization profile chosen in Eq.~(\ref{eq:magnetization}) may come across as ad hoc, even if we found that it is more effective in inducing snap buckling than the a uniform magnetization. Future work should explore other magnetization designs more systematically. Due to the complex relationship between the design parameters and snap-through characteristics, modeling the deformation of bistable beams under other boundary conditions should be considered.

It is important to highlight that, a recent study on hard-magnetic plates~\cite{yan2022reduced} proposed a rotation-based ($\mathbf{R}$-based) magnetic potential by replacing the deformation gradient, $\mathbf{F}$, in Eq.~\eqref{Um_3D_beam} with the rotation tensor. This work was, in turn, motivated by an also equally recent, but prior, demonstration of the stretch-independence of the magnetization of bulk h-MREs~\cite{mukherjee2021explicit}. Indeed, the subsequent experiments in Ref.~\cite{yan2022reduced} showed that the $\mathbf{R}$-based model is necessary for plates subjected to non-negligible stretching deformation under an applied field parallel to the initial magnetization. These latest findings bring into question why, in the present paper, we decided not to use the $\mathbf{R}$-based magnetic potential, choosing the $\mathbf{F}$-based description instead. Given the assumption of inextensible centerline made when developing the beam model, together with the orthogonality between the field and the initial magnetization of the configuration considered in this work, the potential in Eq.~\eqref{Um_3D_beam} is expected to be appropriate for the current problem, as also justified by the excellent agreement between our theory, FEM, and experiments. Also, the Kirchhoff assumptions adopted in the 1D model correct the error from using the potential in Eq.~\eqref{Um_3D_beam}, as pointed out in Ref.~\cite{yan2022reduced}, given that inextensibility together with the fact that normals do not change length, thereby removing any stretching-induced effects from the $\mathbf{F}$-based. As a final practical justification, the $\mathbf{F}$-based being  significantly simpler mathematically than the $\mathbf{R}$-based one and, therefore, it is preferable in cases where both yield the same results. Still, future efforts should be dedicated to develop and $\mathbf{R}$-based beam and rod models for more general cases where stretching of the centerline may be important.

In closing, we believe that our comprehensive framework is a step forward toward the predictive design of bistable magneto-elastic beams. We hope that the snapping behavior and the stiffness-tuning capability of these components will be exploited for a variety of future applications, including actuators, robotics, MEMS, programmable devices, metamaterials, and energy harvesting devices.
\bigskip

\noindent \textbf{Acknowledgments:} A.A. is grateful to the support from the Federal Commission for Scholarships for Foreign Students (FCS) through a Swiss Government Excellence Scholarship (Grant No.  2019.0619). 
\appendix
\label{sec:appendixA}
\section{Solution of Eq.~(\ref{simpODE})}
\label{sec:Var Parameters}

Here, we discuss a few details of the solution of Eq.~(\ref{simpODE}) to arrive at Eq.~(\ref{eq:lin_sol}).
The functions, $\varphi_1$ and $\varphi_2$, were introduced to solve the linear inhomogeneous ODE with the method of variation of parameters (Section~\ref{sec:Beam_mechmag}) and are written as 
\begin{eqnarray}
\varphi_{1}(s) \equiv 1-\frac{\cos \left(\kappa\left(s-\frac{1}{2}\right)\right)}{\cos (\kappa / 2)},
\end{eqnarray}

\begin{equation}
\varphi_{2}(s) \equiv 
\frac{1}{2}\left[{\rm sgn}\left(s-\frac{1}{2}\right)\left\{\cos\left(\kappa\left(s-\frac{1}{2}\right)\right)-1\right\} + \tan\left(\frac{\kappa}{4}\right)\sin\left(\kappa\left(s-\frac{1}{2}\right)\right)\right],
\end{equation}
where we have used $\tan(\kappa/4)=(1-\cos(\kappa/2))/\sin(\kappa/2)$.
Note that $\varphi_1(s)$ and $\varphi_2(s)$ are, respectively, symmetric and asymmetric function with respect to ${s}=1/2$.  
The two numerical constants in Eq.~(\ref{ode}), $c_1$ and $c_2$, are computed as
\begin{equation}
\begin{aligned}
c_{1} &\equiv\int_{0}^{1} \varphi_{1}^{2}(s) d s
=\frac{2 \kappa-3 \sin \kappa+\kappa \cos \kappa}{\kappa+\kappa \cos \kappa},\\
c_{2} &\equiv\int_{0}^{1} \varphi_{2}^{2}(s) d s
=\frac{1}{8}\left(2 + \frac{1}{\cos^2(\kappa/4)} - \frac{12}{\kappa}\tan\left(\frac{\kappa}{4}\right)\right).
\end{aligned}
\end{equation}

\bibliography{Bibliography}

\end{document}